# Two γ-ray bursts from a dust-rich regions with little molecular gas

B. Hatsukade[1], K. Ohta[2], A. Endo[3], K. Nakanishi[1,4,5], Y. Tamura[6], T. Hashimoto[1], & K. Kohno[6,7]

[1]National Astronomical Observatory of Japan, 2-21-1 Osawa, Mitaka, Tokyo 181-8588, Japan.
[2]Department of Astronomy, Kyoto University, Kyoto 606-8502, Japan.
[3]Kavli Institute of NanoScience, Faculty of Applied Sciences, Delft University of Technology, Lorentzweg 1, 2628 CJ Delft, The Netherlands
[4]Joint ALMA Observatory, Alonso de Córdova 3107, Vitacura, Santiago 763 0355, Chile
[5]The Graduate University for Advanced Studies (SOKENDAI), 2-21-1 Osawa, Mitaka, Tokyo 181-8588, Japan
[6]Institute of Astronomy, University of Tokyo, 2-21-1 Osawa, Mitaka, Tokyo 181-0015, Japan
[7]Research Centre for the Early Universe, University of Tokyo, 7-3-1 Hongo, Bunkyo, Tokyo 113-0033, Japan

**Long-duration γ-ray bursts are associated with the explosions of massive stars[1] and are accordingly expected to reside in star-forming regions with molecular gas (the fuel for star formation). Previous searches for carbon monoxide (CO), a tracer of molecular gas, in burst host galaxies did not detect any emission[2-4]. Molecules have been detected as absorption in the spectra of γ-ray burst afterglows, and the molecular gas is similar to the translucent or diffuse molecular clouds of the Milky Way[5,6]. Absorption lines probe the interstellar medium only along the line of sight, so it is not clear whether the molecular gas represents the general properties of the regions where the bursts occur. Here we report spatially resolved observations of CO line emission and millimetre-wavelength continuum emission in two galaxies hosting γ-ray bursts. The bursts happened in regions rich in dust, but not particularly rich in molecular gas. The ratio of molecular gas to dust (<9-14) is significantly lower than in star-forming regions of the Milky Way and nearby star-forming galaxies, suggesting that much of the dense gas where stars form has been dissipated by other massive stars.**

We selected the two γ-ray burst (GRB) hosts (GRB 020819B at a redshift of $z = 0.41$ and GRB 051022 at $z = 0.81$) with high star formation rates (SFRs) and high gas metallicity among GRB host galaxies to maximize the possibility of detecting the CO emission line and dust continuum emission. The GRB 020819B host shows an extinction-corrected SFR of ~10–30 $M_\odot$ yr$^{-1}$ (where $M_\odot$ indicates solar mass) derived from ultraviolet continuum emission, the Hα emission line, and spectral energy distribution (SED) fitted using infrared data[7-9]. The SFRs at spatially resolved positions are also derived from the Hα emission line, which are 10.2 $M_\odot$ yr$^{-1}$ and 23.6 $M_\odot$ yr$^{-1}$ at the nuclear region and at the GRB explosion site, respectively[8]. The host galaxy of GRB 051022 shows an extinction-corrected SFR of ~20–70 $M_\odot$ yr$^{-1}$ derived from ultraviolet continuum emission, the [O II] emission line at rest-frame wavelength λ = 3727 Å, the SED fitted with infrared data, and radio continuum emission[7,9-11]. The gas metallicity is measured at the GRB 020819B site, the nuclear region of the GRB 020819B host, and at the GRB 051022 host, and they all have at least solar metallicity[8,12]. The two GRBs are classified as `dark GRBs'[13,14], whose afterglow is optically dark compared with what is expected from X-ray afterglows[15]. The origin of the dark GRBs is not yet well understood, but a plausible explanation is dust obscuration along the line of sight to GRBs[16-19].

We conducted observations of the CO emission line and 1.2-mm continuum towards the GRB hosts using the Atacama Large Millimeter/submillimeter Array (ALMA). We observed the redshifted CO(3–2) line for the GRB 020819B host and the CO(4–3) line for the GRB 051022 host. The angular resolution is ~0.8″ × 0.7″ (~4 kpc × 4 kpc) and ~1.0″ × 0.7″ (~8 kpc × 5 kpc) (full-width at half maximum; FWHM) for the GRB 020819B host and the GRB 051022 host, respectively. The GRB 020819B host is spatially resolved in the observations. The CO emission line is clearly detected at the nuclear region of the GRB 020819B host and the GRB 051022 host (Figs 1a and d and 2). While molecular gas has been detected in absorption in spectra of GRB afterglows[5,6], this is the first case for detecting spatially resolved molecular gas emission in GRB hosts[2-4]. The component size of the CO emission (deconvolved from beam) derived from a Gaussian fitting is 3.2 × 1.5 kpc (FWHM). The molecular gas mass estimated from the CO emission is $M_{gas}$ = 2.4 × 10$^9$ $M_\odot$ and 2.1 × 10$^9$ $M_\odot$ for the nuclear region of the GRB 020819B host and the GRB 051022 host, respectively (see Methods and Table 1). The molecular gas mass is comparable to those of local massive spiral galaxies[20], and lower than those of $z \approx$ 1–2 normal star-forming galaxies[21] or submillimetre-luminous galaxies[22,23]. The fraction of molecular gas mass to stellar mass[7,24] for the hosts is ~0.1, which is comparable to those of local spiral galaxies[20].

The 1.2-mm continuum emission is also detected in both GRB hosts (Fig. 1b, e). The spatially resolved continuum map of the GRB 020819B host shows that the emission is significantly detected only at a star-forming region ~3″ (16 kpc in projection) away from the nuclear region, where the GRB explosion occurred. The size of the continuum emission deconvolved from the telescope beam is ~1.7 kpc × 1.0 kpc. We regard the continuum emission as dust thermal emission originating in star-forming activity (see Supplementary Information). By assuming that the dust emission is described as a modified blackbody and using the dust temperature and emissivity index derived from fitting, we derive the dust mass of $M_{dust}$ = 4.8 × 10$^7$ $M_\odot$ and 2.9 × 10$^7$ $M_\odot$ for the GRB 020819B site and the GRB 051022 host, respectively (see Methods). The far-infrared luminosity and SFR are $L_{FIR}$ = 1.1 × 10$^{11}$ $L_\odot$ (where $L_\odot$ is solar luminosity) and SFR = 18 $M_\odot$ yr$^{-1}$ for the GRB 020819B site, and $L_{FIR}$ = 1.9 × 10$^{11}$ $L_\odot$ and SFR = 32 $M_\odot$ yr$^{-1}$ for the GRB 051022 host, respectively. The SFRs are comparable to the extinction-corrected SFRs derived from ultraviolet and optical observations, suggesting that there is no sign of an extra, optically completely invisible portion of star formation that cannot be recovered by extinction correction.

Of particular interest is that the spatial distributions of molecular gas and dust are clearly different in the GRB 020819B host. The ratio of molecular gas mass to dust mass of the GRB 020819B host is >51–60 and <9–14 (3σ limits with uncertainty from dust mass) at the nuclear region and the GRB site, respectively. The ratio in the GRB site is significantly lower than that of the nuclear region, indicating that the GRB occurred under particular circumstances within the host. The molecular gas-to-dust ratio at the GRB site is also lower than those of the Milky Way and nearby star-forming galaxies[25], suggesting that the star-forming environment where GRBs occur is different from those in local galaxies. While the correlation between gas-to-dust ratio and metallicity has been observed[20,25], it is unlikely that the large difference between the GRB site and the nuclear region is attributable to the difference in metallicity because both regions have a similar metallicity[8]. The difference of distribution between molecular gas and dust is also seen in the GRB 051022 host and the GRB site seems to be a dust-rich region, although the angular resolution is not good enough to be certain. The possible reasons for the deficit of molecular gas in the GRB site are that much of the dense gas where stars form has been incorporated into stars, or dissipated by a strong interstellar ultraviolet radiation field, which is expected in regions with intense star formation. The lack of molecular gas in optical spectra of GRB afterglows has been reported[26] and a possible explanation is the dissociation of molecules by ambient ultraviolet radiation with 10–100 times the Galactic mean value from the star-forming regions where GRB progenitors reside[26,27]. GRB hosts with a mean ultraviolet radiation field of 35–350 times the Galactic mean value have been observed[28]. The molecular gas-to-dust ratio in GRB hosts could be an important indicator of an environment where GRBs occur.

The occurrence of GRB 020819B in a dust-rich region supports the idea that the dust extinction is the cause for the darkness of optical afterglow[13,24]. The molecular gas-to-dust ratio in the GRB site is comparable to or lower than the ratios in nuclear regions of local galaxies of ~20–40 (ref. 29) and submillimetre-luminous galaxies of ~50 (ref. 30), suggesting the existence of GRBs that could occur in dusty galaxies such as submillimetre-luminous galaxies.

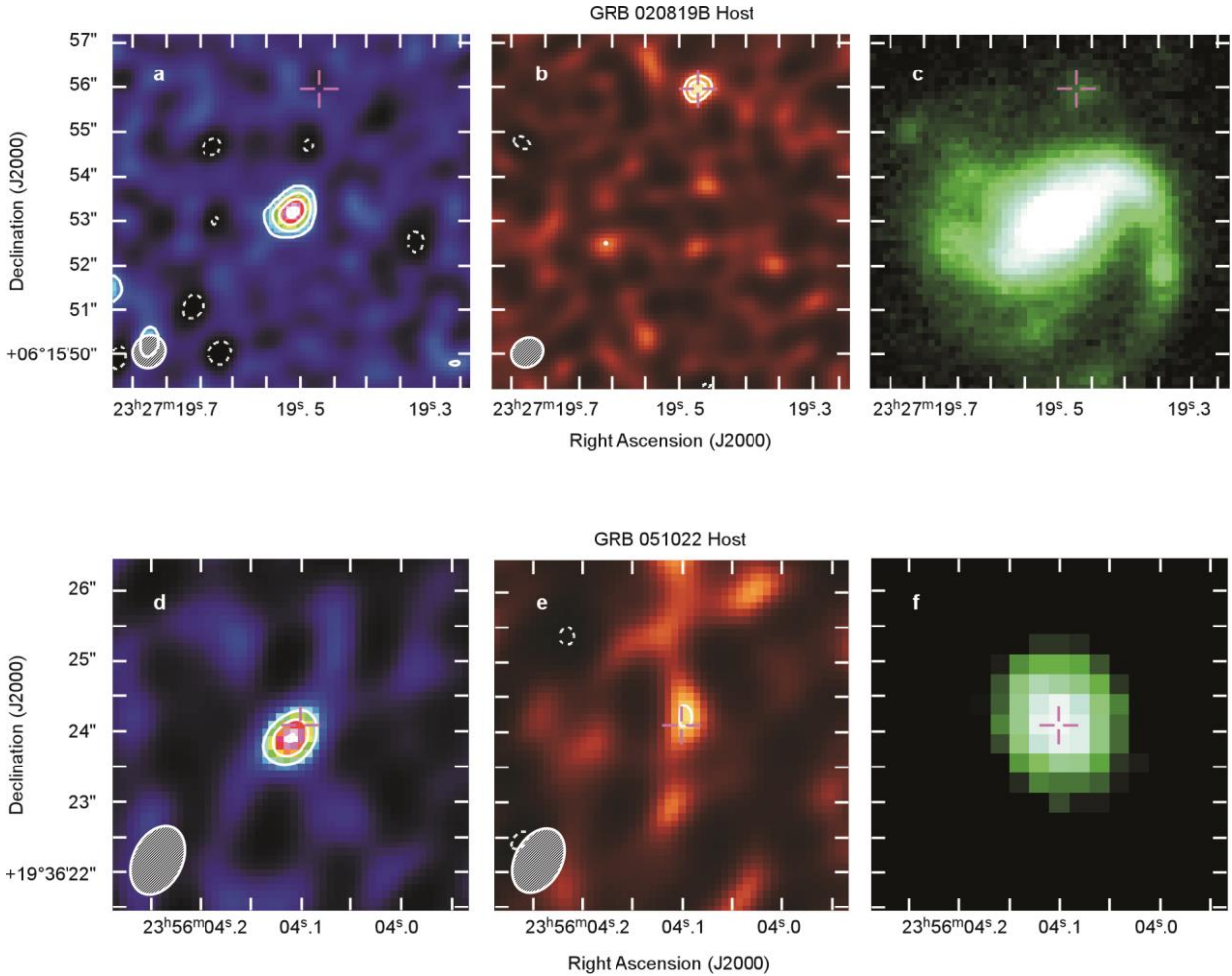

**Figure 1 | CO maps, 1.2-mm continuum maps, and optical images of the GRB hosts.** The magenta cross represents the position of the radio afterglow. ALMA beam size is shown in the lower left corners of **a**, **b**, **d**, and **e** (black and white ovals). **a**, Velocity-integrated CO(3–2) intensity map. Contours start from ±3σ with 2σ step (1σ = 0.040 Jy beam$^{-1}$ km s$^{-1}$). **b**, 1.2-mm continuum map. Contours start from ±3σ with 1σ step (1σ = 0.030 mJy beam$^{-1}$). **c**, Optical *R*-band image obtained with the Gemini North Telescope. **d**, Velocity-integrated CO(4–3) intensity map. Contours start from 3σ with 1σ step (1σ = 0.037 Jy beam$^{-1}$ km s$^{-1}$). **e**, 1.2-mm continuum. Contours are ±3σ (1σ = 0.032 mJy beam$^{-1}$). **f**, Optical *R*-band image obtained with the Gemini South Telescope.

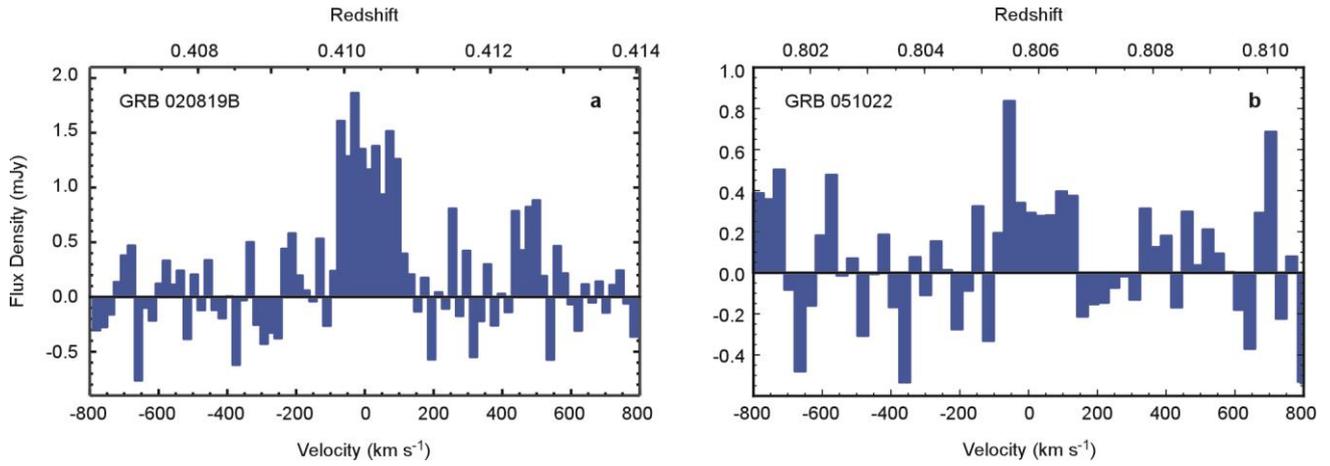

**Figure 2 | CO spectra of the GRB hosts.** Continuum emission is subtracted. **a**, CO(3–2) spectrum of the nuclear region of the GRB 020819B host at 20 km s$^{-1}$ resolution. A Gaussian fit to the emission line gives a redshift of $z = 0.410$ and a velocity width of 167 km s$^{-1}$ (FWHM). **b**, CO(4–3) spectrum of the GRB 051022 host at 30 km s$^{-1}$ resolution. A Gaussian fit to the emission line gives a redshift of $z = 0.806$ and a velocity width of 176 km s$^{-1}$ (FWHM).

**Table 1 | Properties of GRB host galaxies**

|  | GRB 020819B nuclear region | GRB 020819B site | GRB051022 |
|---|---|---|---|
| $z_{CO}$ | 0.410 |  | 0.806 |
| CO transition | 3–2 |  | 4–3 |
| $L'_{CO(1-0)}$ (K km s$^{-1}$ pc$^2$) | $(5.5 \pm 0.4) \times 10^8$ | $<1.3 \times 10^8$ | $(4.9 \pm 0.9) \times 10^9$ |
| $M_{gas}$ ($M_\odot$) | $(2.4 \pm 0.2) \times 10^9$ | $<5.4 \times 10^8$ | $(2.1 \pm 0.4) \times 10^9$ |
| $S_{1.2\,mm}$ (mJy) | $<0.12$ | $0.14 \pm 0.03$ | $0.10 \pm 0.03$ |
| $M_{dust}$ ($M_\odot$) | $<4.2 \times 10^7$ | $(4.8 \pm 1.0) \times 10^7$ | $(2.9 \pm 0.9) \times 10^7$ |
| $L_{FIR}$ ($L_\odot$) | $<9.3 \times 10^{10}$ | $(1.1 \pm 0.2) \times 10^{11}$ | $(1.9 \pm 0.6) \times 10^{11}$ |
| SFR ($M_\odot$ yr$^{-1}$) | $<16$ | $18 \pm 4$ | $32 \pm 10$ |
| $M_{gas}/M_{dust}$ | $>51$–$60$ | $<9$–$14$ | $58$–$86$ |

The errors represent root-mean-square (1σ) uncertainties from the photometry error. The limits are 3σ. We adopt a cosmology with $H_0 = 71$ km s$^{-1}$ Mpc$^{-1}$, $\Omega_M = 0.27$, and $\Omega_\Lambda = 0.73$. For details, see Methods. $z_{CO}$ is the redshift determined from the CO line. $L'_{CO(1-0)}$ is the CO(1–0) luminosity derived from $L'_{CO} = 3.25 \times 10^7 S_{CO}\Delta v \nu_{obs}^{-2} D_L^2 (1+z)^{-3}$, where $L'_{CO}$ in units of K km s$^{-1}$ pc$^2$, $S_{CO}\Delta v$ is the velocity-integrated flux in Jy km s$^{-1}$, $\nu_{obs}$ is the observed line frequency in GHz, $D_L$ is the luminosity distance in Mpc. We assume a CO line ratio of CO(3–2)/CO(1–0) = 0.93 and CO(4–3)/CO(1–0) = 0.85, which are the values for the local star-forming galaxy M82. $M_{gas}$ is the molecular gas mass derived from $M_{gas} = \alpha_{CO} L'_{CO(1-0)}$, where $\alpha_{CO}$ is the CO-to-molecular gas mass conversion factor of a Galactic value $\alpha_{CO} = 4.3$ in units of $M_\odot$ (K km s$^{-1}$ pc$^2$)$^{-1}$. $S_{1.2\,mm}$ is the 1.2-mm continuum flux. $M_{dust}$ is the dust mass derived from $M_{dust} = S_{obs} D_L^2 / [(1+z) \kappa_d(\nu_{rest}) B(\nu_{rest}, T_d)]$, where $S_{obs}$ is the observed flux density, $\nu_{rest}$ is the rest frequency, $\kappa_d(\nu_{rest})$ is the rest-frequency mass absorption coefficient, $B(\nu_{rest}, T_d)$ is the Planck function. $L_{FIR}$ is the far-infrared luminosity derived from $L_{FIR} = 4\pi M_d \int_0^\infty \kappa_d(\nu) B(\nu, T_d) d\nu$. SFR is the star-formation rate derived from SFR (in $M_\odot$ yr$^{-1}$) = $1.72 \times 10^{-10} L_{FIR}$ (in $L_\odot$).

# METHODS SUMMARY

We conducted ALMA observations of the GRB 020819B host and the GRB 051022 host at 245.072 GHz and 255.142 GHz, respectively, with a bandwidth of 1,875 MHz and with 24-27 antennas. The data were reduced with the Common Astronomy Software Applications package in a standard manner. The maps were processed with the CLEAN algorithm with Briggs weighting (with the robust parameter equal to 0.5). The final synthesized beam size (FWHM) is ~0.8″ × 0.7″ and ~1.0″ × 0.7″ for the GRB 020819B host and the GRB 051022 host, respectively. We derived the molecular gas mass of $M_{gas} = (2.4 \pm 0.2) \times 10^9\ M_\odot$ and $(2.1 \pm 0.4) \times 10^9\ M_\odot$ for the nuclear region of the GRB 020819B host and the GRB 051022 host, respectively. Here we assume CO line ratios of CO(3–2)/CO(1–0) = 0.93 and CO(4–3)/CO(1–0) = 0.85, which are the values for the local star-forming galaxy M82, by considering the star-forming property of the hosts. We adopt the CO-to-molecular gas mass conversion factor of Galactic value (4.3 $M_\odot$ (K km s$^{-1}$ pc$^2$)$^{-1}$) because the metallicity of the two hosts is close to the solar metallicity. To estimate a dust temperature and an emissivity index, we fitted a single temperature modified blackbody form to the far-infrared-millimetre photometry of the ALMA 1.2-mm data and Herschel Space Observatory 100-μm, 160-μm, and 250-μm data. The best-fitting results are $T_{dust} = 28 \pm 3$ K and $\beta = 1.9 \pm 0.3$ for the GRB 020819B host and $T_{dust} = 34 \pm 6$ K and $\beta = 1.8 \pm 0.5$ for the GRB 051022 host. By using the best-fit modified blackbody functions, we derived dust mass of $(4.8 \pm 1.0) \times 10^7\ M_\odot$ and $(2.9 \pm 0.9) \times 10^7\ M_\odot$ for the GRB 020819B site and the GRB 051022 host, respectively.

**Acknowledgements**

We acknowledge Akiko Kawamura and ALMA staffs for helpful support. We are grateful to Kiyoto Yabe and Akifumi Seko for useful discussions. We thank the editor and referees for helpful comments and suggestions. B.H. was supported by a Japan Society for the Promotion of Science (JSPS) Grant-in-Aid for JSPS Fellows. K.O. was supported by a JSPS Grant-in-Aid for Scientific Research (C) (grant number 24540230). A.E. was supported by the NWO (Veni grant number 639.041.023). Y.T. was supported by JSPS Grant-in-Aid for Scientific Research on Innovative Areas (grant number 25103503). ALMA is a partnership of ESO (representing its member states), NSF (USA) and NINS (Japan), together with NRC (Canada) and NSC and ASIAA (Taiwan), in cooperation with the Republic of Chile. The Joint ALMA Observatory is operated by ESO, AUI/NRAO and NAOJ. This research is based in part on observations made with Herschel. Herschel is an ESA space observatory with science instruments provided by European-led Principal Investigator consortia and with important participation from NASA. This research makes use of data based on observations obtained at the Gemini Observatory, which is operated by the Association of Universities for Research in Astronomy, Inc., under a cooperative agreement with the NSF on behalf of the Gemini partnership: the National Science Foundation (United States), the National Research Council (Canada), CONICYT (Chile), the Australian Research Council (Australia), Ministério da Ciência, Tecnologia e Inovação (Brazil) and Ministerio de Ciencia, Tecnología e Innovación Productiva (Argentina).

**Author Contributions**

B.H. had the overall lead of the project. B.H. reduced the ALMA data and wrote the manuscript. K.O. conducted the photometry of the Gemini and Herschel data. All authors contributed to the ALMA proposal, discussed the results and implications, and commented on the manuscript.

**Author Information**

This research is based on the following ALMA data: ADS/JAO.ALMA#2011.0.00232.S. Reprints and permissions information is available at www.nature.com/reprints. The authors declare no competing financial interests. Correspondence and requests for materials should be addressed to B.H. (bunyo.hatsukade@nao.ac.jp)

## METHODS

**Observations, data reduction, and results.** We conducted ALMA band-6 observations of the GRB 020819B host in 2012 November 17 with 27 antennas and the GRB 051022 host in 2012 November 21 and December 2 with 24 antennas during the ALMA cycle-0 session. The range of baseline lengths of the configuration is 15-402 m and 15-382 m for the observations of the GRB 020819B host and the GRB 051022 host, respectively. The maximum recoverable scale (the largest angular structure to which a given array is sensitive) for the array configurations is 10", which is large enough to cover the angular scale of the host galaxies. The correlator was used in the frequency domain mode with a bandwidth of 1,875 MHz (488.28 kHz × 3,840 channels). Four basebands were used, giving a total bandwidth of 7.5 GHz. We observed the redshifted CO(3–2) line at 245.072 GHz for the GRB 020819B host and the redshifted CO(4–3) line at 255.142 GHz for the GRB 051022 host. Uranus was observed as a flux calibrator and a quasar J2253+161 (3C454.3) was observed for bandpass and phase calibrations. The on-source time is 47 min and 71 min for the GRB 020819B host and the GRB 051022 host, respectively. The data were reduced with the Common Astronomy Software Applications[31] package in a standard manner. The maps were processed with the CLEAN algorithm with Briggs weighting (with robust parameter of 0.5). The final synthesized beam size (FWHM) is ~0.8″ × 0.7″ and ~1.0″ × 0.7″ for the GRB 020819B host and the GRB 051022 host, respectively. 1.2-mm continuum maps were created with a total bandwidth of ~7.5 GHz, excluding channels with emission line. CO emission and 1.2-mm continuum emission are detected at both GRB host galaxies (Figs 1a, b, d and e, and 2). The GRB 020819B host is spatially resolved in the observations, while the GRB 051022 host is not. The velocity-integrated CO intensity is $S_{CO(3-2)} = 0.53 \pm 0.04$ Jy km s$^{-1}$ and $S_{CO(4-3)} = 0.19 \pm 0.03$ Jy km s$^{-1}$ at the nucleus of the GRB 020819B host and the GRB 051022 host, respectively. The 1.2-mm continuum flux density is $S_{1.2mm} = 0.14 \pm 0.03$ mJy and $S_{1.2mm} = 0.10 \pm 0.03$ mJy at the explosion site of GRB 020819B and the GRB 051022 host, respectively.

**Molecular Gas Mass.** CO luminosity is derived from $L'_{CO} = 3.25 \times 10^7 S_{CO}\Delta v \, \nu_{obs}^{-2} D_L^2 (1+z)^{-3}$ (ref. 23), where $L'_{CO}$ in units of K km s$^{-1}$ pc$^2$, $S_{CO}\Delta v$ is the velocity-integrated flux in Jy km s$^{-1}$, $\nu_{obs}$ is the observed line frequency in GHz, and $D_L$ is the luminosity distance in Mpc. We assume a CO line ratio of CO(3–2)/CO(1–0) = 0.93 and CO(4–3)/CO(1–0) = 0.85, which are the values for the local star-forming galaxy M82 (ref. 32), by considering the star-forming property of the host galaxies. The derived CO(1–0) luminosity is $(5.5 \pm 0.4) \times 10^8$ (K km s$^{-1}$ pc$^2$) and $(4.9 \pm 0.9) \times 10^9$ (K km s$^{-1}$ pc$^2$) for the nuclear region of the GRB 020819B host and the GRB 051022 host, respectively. Molecular gas mass is derived from $M_{gas} = \alpha_{CO} L'_{CO(1-0)}$, where $\alpha_{CO}$ is the CO-to-molecular gas mass conversion factor in units of $M_\odot$ (K km s$^{-1}$ pc$^2$)$^{-1}$ including He mass. It is thought that there is a correlation between $\alpha_{CO}$ and metallicity in the local Universe and at $z \approx 1–2$ (refs. 33, 34); $\alpha_{CO}$ decreases with increasing metallicity. Because the metallicity of the two hosts is close to the solar metallicity, we adopt the Galactic value of $\alpha_{CO} = 4.3 \, M_\odot$ (K km s$^{-1}$ pc$^2$)$^{-1}$ (ref. 35). The derived molecular gas masses are $M_{gas} = (2.4 \pm 0.2) \times 10^9 \, M_\odot$ and $(2.1 \pm 0.4) \times 10^9 \, M_\odot$ for the nuclear region of the GRB 020819B host and the

GRB 051022 host, respectively.

**Photometry of Herschel Space Observatory[36] data.** We used the Herschel/Photodetector Array Camera and Spectrometer (PACS)[37] data in the archive. We conducted aperture photometry on the 160-μm image of the GRB 051022 host with SExtractor[38] and obtained a flux density of $S_{160\ \mu m}$ = 12 mJy (with about 30% photometry error). There is no significant contamination from nearby sources to the photometry. The FWHM of the source size is ~14″ at 160 μm, respectively, which is comparable to the FWHM of PACS beam size[37]. We also measured the centroid of the 100-μm emission of the GRB 020819B host and found that the emission is in between the galaxy centre and the peak of 1.2-mm continuum. It is possible that dust is more widely spread in the host galaxy, although the angular resolution is inadequate (FHWM of ~7″).

**Modified blackbody fit.** To estimate a dust temperature ($T_{\rm dust}$) and an emissivity index ($\beta$), we fitted the far-infrared-millimetre photometry data of Herschel 100-μm, 160-μm, and 250-μm (ref. 9), and ALMA 1.2 mm with a single temperature modified blackbody form of $S_\nu \propto \nu^{3+\beta}/(\exp(h\nu/kT_{\rm dust})-1)$, where $S_\nu$ is the flux density and $\nu$ is the frequency (Extended Data Fig. 1). The best-fitting results are $T_{\rm dust}$ = 28 ± 3 K and $\beta$ = 1.9 ± 0.3 for the GRB 020819B host and $T_{\rm dust}$ = 34 ± 6 K and $\beta$ = 1.8 ± 0.5 for the GRB 051022 host. We note that the missing flux in the ALMA observations in the scale of the PACS beamsize is negligible. The dust temperatures are within the typical range of $z \approx$ 0–2 star-forming galaxies[39,40]. The dust temperatures of the hosts were derived in a previous study with a SED model fit to optical-infrared data including Herschel photometry[9]: $T_{\rm dust}$ = 24.4 K and 52.6 K for the GRB 020819B host and the GRB 051022 host, respectively. The dust temperature of the GRB 051022 host is higher than in this work. This may be due to their lack of photometric data at >160 μm, which is essential to fit dust SED.

**Dust mass, far-infrared luminosity, and SFR.** By using the best-fitting modified blackbody functions, we estimated dust mass, far-infrared luminosity, and SFR. Dust mass is derived by $M_{\rm dust}$ = $S_{\rm obs}D_{\rm L}^2/[(1+z)\ \kappa_{\rm d}(\nu_{\rm rest})B(\nu_{\rm rest}, T_{\rm dust})]$ (ref. 411), where $S_{\rm obs}$ is the observed flux density, $\nu_{\rm rest}$ is the rest frequency, $\kappa_{\rm d}(\nu_{\rm rest})$ is the rest frequency mass absorption coefficient, $B(\nu_{\rm rest}, T_{\rm dust})$ is the Planck function. We assume that the absorption coefficient varies as $\kappa_{\rm d}(\nu) \propto \nu^\beta$ and $\kappa_{\rm d}(125\ \mu m)$ = 26.4 cm$^2$ g$^{-1}$ (ref. 422). The derived dust mass is (4.8 ± 1.0) × 10$^7$ $M_\odot$ and (2.9 ± 0.9) × 10$^7$ $M_\odot$ for the GRB 020819B site and the GRB 051022 host, respectively. If we use the dust temperature of 52.6 K for the GRB 051022 host estimated in a previous work[9], the derived dust mass would be about a factor of two lower, which has no effect on the discussion in the main text. Far-infrared luminosity is derived from $L_{\rm FIR}$ = $4\pi M_{\rm dust} \int_0^\infty \kappa_{\rm d}(\nu)B(\nu, T_{\rm dust})d\nu$ (ref. 411). The derived far-infrared luminosity is (1.1 ± 0.2) × 10$^{11}$ $L_\odot$ and (1.9 ± 0.6) × 10$^{11}$ $L_\odot$ for the GRB 020819B site and the GRB 051022 host, respectively. SFR is derived from the far-infrared luminosity as follows: SFR (in $M_\odot$ yr$^{-1}$) = 1.72 × 10$^{-10}$ $L_{\rm FIR}$ (in $L_\odot$) (ref. 433), and calculated to be 18 ± 4 $M_\odot$ yr$^{-1}$ and 32 ± 10 $M_\odot$ yr$^{-1}$ for the GRB 020819B site and the GRB 051022 host, respectively. The comparison of CO and far-infrared luminosities is shown in Extended Data Fig. 2.

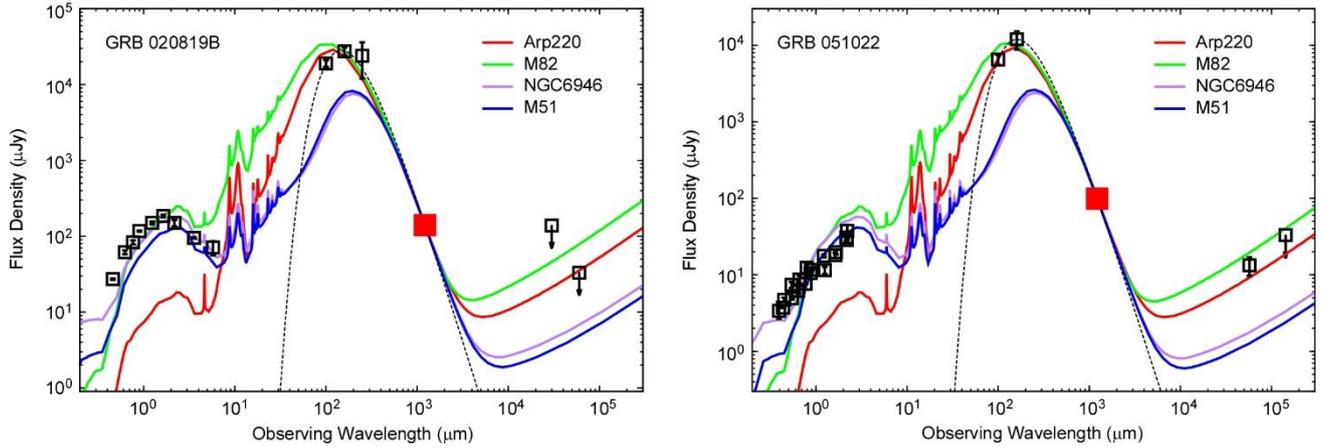

**Extended Data Figure 1 | Spectral energy distribution of GRB 020819B host and GRB 051022 host.** The red squares show ALMA 1.2-mm data. Black squares represent photometry from literatures[10,11,14,24,44,45,466] and the publicly archived data. Dashed curves show the best-fit modified blackbody functions. The arrows represent 3σ upper limits. For comparison, we plot SED models of Arp220, M82, NGC6946, and M51 (ref. 47). The SED models are scaled to the flux density of ALMA data.

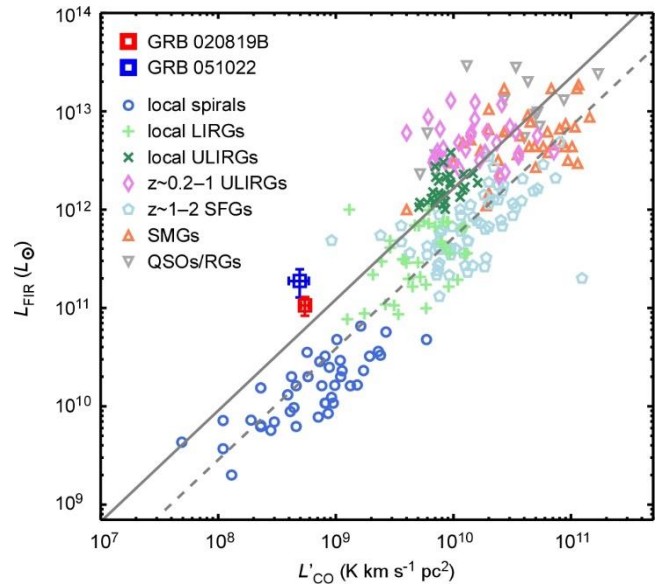

**Extended Data Figure 2 | Comparison of CO and far-infrared luminosities.** The GRB 020819B host and the GRB 051022 host are plotted with 1σ uncertainties (red and blue squares). To examine the properties of the GRBhost galaxies as a whole and to compare with previous studies, we plot our data without separating the nuclear region and the explosion site for the GRB020819B host galaxy. Various galaxy populations are also plotted: local spirals[20,48] (circles), local LIRGs (plus symbols) and ultraluminous infrared galaxies (ULIRGs)[23,48] (crosses), $z \approx 0.2$–1 ULIRGs[49,500] (diamonds), $z \approx 1$–2 normal star-forming galaxies[21] (pentagons), submillimetre-luminous galaxies[22,23] (up-triangles), QSOs and radio galaxies[23] (down-triangles). The grey solid and dashed lines represent the sequence of normal star-forming galaxies and starburst galaxies, respectively[51].

# SUPPLEMENTARY INFORMATION

**Spectral Energy Distribution**

We created spectral energy distributions (SEDs) of the host galaxies of GRB 020819B and GRB 051022 from optical to radio wavelength by using photometry data of our ALMA observations, the archive data of *Herschel*, and literatures. We plot the SEDs of the GRB 020819B host and GRB 051022 host in Extended Data Fig. 1. For comparison, we show the SED models[47] of local galaxies Arp220 (archetypal ultra-luminous infrared galaxy; ULIRG), M82 (prototype starburst), NGC6946, and M51 (spiral galaxy). The SED models are redshifted for each host galaxy and normalized to the flux density of ALMA observations. The SEDs of both GRB hosts are closer to those of starburst galaxies than those of spiral galaxies.

**Origin of the Continuum Emission in the GRB 020819B Host Galaxy**

The 1.2-mm continuum emission in the GRB 020819B host galaxy is only detected at the GRB site. A similar case is reported in the host galaxy of GRB 980425, where a star-forming region ~800 pc away from the GRB site dominates the IR emission of the entire host[52-54]. The ALMA 1.2-mm photometry data is well explained by the modified blackbody spectrum (Extended Data Fig. 1) and the inferred SFR is comparable to the extinction-corrected SFR derived from ultraviolet and optical observations. These suggest that the origin of the continuum emission at the GRB site is most likely the thermal emission from dust, heated by star-forming activity in the host galaxy. Nevertheless, we discuss alternative possibilities for the origin of the 1.2-mm emission below: (1) emission from another galaxy behind the host, and (2) synchrotron emission from the GRB remnant.

(1) The peak position of the 1.2-mm continuum emission is $\alpha(J2000.0) = 23^h27^m19^s.47$ and $\delta(J2000.0) = 06°15'55.95''$. The positional uncertainty for a SN = 4.7 source is ~0.17″. The 1.4 GHz radio afterglow position is $\alpha(J2000) = 23^h27^m19^s.475$ and $\delta(J2000) = 06°15'55.95''$, with an error of 0.5″ (ref. 13). The positional offset between them is 0.06″, which is sufficiently small compared to the positional uncertainty. We estimate the probability that another source behind the host galaxy is coincidentally detected at the GRB position. The spatial density of sources with $S_{1.2mm} \geq 0.14$ is $\sim 3 \times 10^4$ deg$^{-2}$ based on the recently obtained number counts at 1.3 mm (ref. 55). The expected number of sources which fall within a radius of 0.06″ is $\sim 3 \times 10^{-5}$. The chance of coincidence is negligible and we can conclude that the 1.2-mm continuum emission comes from the GRB site.

(2) The radio continuum observations of the GRB 020819B host at 3 cm and 6 cm in January 2010 did not detect emission, giving the 3σ upper limits of $S_{3cm} < 138$ μJy and $S_{6cm} < 33$ μJy (ref. 45). Because the spectral index of synchrotron emission against the wavelength is positive, the upper limits reject the possibility that the 1.2-mm continuum emission is attributed to synchrotron emission.

Considering the above, we can regard the continuum emission as the thermal emission from dust, heated by star-forming activity. We note that there is a possibility that the dust was heated by the GRB explosion. However, we do not consider this possibility here because many uncertainties remain (e.g., the amount of dust, the geometry of dust, how much of the explosion energy is transferred to dust heating).

**Comparison of CO and Far-infrared Luminosities**

Recent molecular gas observations of local and high-redshift star-forming galaxies suggest the existence of two different star-formation modes[51]: (1) a long-lasting mode for disks seen in local spirals and $z \sim 1$–2 normal star-forming galaxies, and (2) a more rapid mode for starburst galaxies seen in local ULIRGs and submillimetre-luminous galaxies. It is proposed that the two types of galaxies are located along separate sequences on the $L'_{CO}$–$L_{FIR}$ plane (Extended Data Fig. 2). In order to examine the properties of the GRB host galaxies as a whole and to compare with previous studies, we plot our data without separating the nuclear region and the explosion site for the GRB 020819B host. The two GRB hosts are closer to the sequence for starburst galaxies, suggesting that the star-forming activity in the hosts has the property of starburst galaxies.